\documentstyle[aps,preprint,epsf]{revtex}
\begin{document}
\draft

\newcommand{\pp}[1]{\phantom{#1}}
\newcommand{\be}{\begin{eqnarray}}
\newcommand{\ee}{\end{eqnarray}}
\newcommand{\ve}{\varepsilon}
\newcommand{\vs}{\varsigma}
\newcommand{\Tr}{{\,\rm Tr\,}}
\newcommand{\pol}{\frac{1}{2}}
\newcommand{\ba}{\begin{array}}
\newcommand{\ea}{\end{array}}
\newcommand{\bear}{\begin{eqnarray}}
\newcommand{\eear}{\end{eqnarray}}
\title{
Correlation experiments in nonlinear quantum mechanics 
}
\author{Marek~Czachor$^{1,2}$ and H.-D. Doebner$^{2}$}
\address{
$^1$ Katedra Fizyki Teoretycznej i Metod Matematycznych\\
Politechnika Gda\'{n}ska,
ul. Narutowicza 11/12, 80-952 Gda\'{n}sk, Poland\\
$^2$ Department of Physics, Technische Universit\"at Clausthal\\
38678 Clausthal-Zellerfeld, Germany\\
e-mail: mczachor@pg.gda.pl (M. Czachor)}
\maketitle

\begin{abstract}
We show how one can compute multiple-time multi-particle 
correlation functions in nonlinear quantum mechanics 
in a way which guarantees locality of the formalism. 
\end{abstract}


\section{Introduction}

Exactly linear dynamics of states is a rarity in physics. 
Linear theories
are in general approximations to nonlinear ones. The exception is quantum
mechanics. Hence the question: Can one construct a consistent
nonlinear theory which contains quantum mechanics as a special case?

Nonlinear extensions of quantum mechanics are not obvious. Nonlinear
Schr\"odinger and von Neumann equations can be justified but do 
not seem to be allowed in the
usual interpretation. A probability interpretation of nonlinear
operators is not clear. 
Of particular interest are difficulties with multi-particle entangled
states
\cite{HB,Gisin,Polchinski,MCu,Jordan,GS,Lucke,MC97,MCMK,MC98,Mielnik}.  
Standard textbook calculations of correlation experiments based on the
projection postulate lead to nonlocal effects. 

In this paper \cite{CD} we
show how to compute conditional and joint probabilities in a way
which eliminates unphysical influences between {\it separated\/} systems, 
and which coincides with the usual prescription if the dynamics is linear. 
We do not exclude the possibility that the scheme we propose should be
modified if 
there exist causal influences between the measurements. Suggestions
concerning a generalization of our proposal can be found
in \cite{G-S}. 

We discuss a two particle system in the tensor space $H_1\otimes H_1$
of one particle Hilbert spaces.  We analyse correlation experiments: 
Measure on particle $\#1$ at time $t_1$ an observable $X_1$ with two possible
outcomes ($+$ or $-$) through a projection operator $E_1$, and on
particle $\# 2$
at time $t_2$ an observable $X_2$ through a projection operator
$E_2$. A nonlinear evolution of the pair of spin-1/2 particles is
constructed via the Polchinski extension from nonlinear one-particle
equations. As opposed to the original Polchinski formulation
\cite{Polchinski} we do not resort to the Many Worlds Interpretation.
It is shown 
that different results are found depending on whether the two
particles are viewed as closed or open systems. In the latter
case there are no nonlocal effects.  In the linear case the two 
alternative approaches give the same result.

The material is arranged as follows.

In section II and III we compare different methods of computing
two-particle correlations in linear quantum mechanics. 
Two approaches are used. 

(a) The two
particles are treated as a closed system. The Hamiltonian is not time
dependent. Two-time probabilities are calculated via
projections-at-a-distance, as used by Gisin \cite{Gisin} and Mielnik
\cite{Mielnik}. 

(b) The two particles are treated as an open system. The
environment contains the measuring devices acting at two different
times. The Hamiltonian is time dependent, the two different times
appearing as parameters. Two-time probabilities are calculated
without reference to projections at-a-distance.
 
Section IV gives a short review of nonlinear evolution equations and
Polchinski's multi-particle extension is introduced. 
Section V contains the central result of this paper: The open-system
generalization of Polchinski's extension. It is shown that 
nonlocal effects do not occur in two-time
measurements if the open-system formalism is employed. 
The results are illustrated in Section VI by explicit solutions for a
two-particle entangled state. 
In Section VII we show how to modify in nonlinear quantum
mechanics the
projection-at-a-distance approach in order to eliminate the nonlocal
effects. 
In Section VIII we discuss problems of probability reduction in
systems which are causally related (preparation at-a-distance,
teleportation, and Russian roulette with a cheating player). 
Some technical points are briefly explained in the Appendix.

\section{Correlation experiments in linear quantum mechanics
--- Heisenberg picture}

For the description of the correlation experiment 
we start with a two-particle entangled state $|\Psi_0\rangle$ 
prepared at time $t=0$. 
The two particles evolve independently by 
unitary operators $V_1(t)=e^{-iH_1 t}$ and 
$V_2(t)=e^{-iH_2 t}$. At times $t_1$
and $t_2$ one performs measurements of two
quantities (``yes-no observables") 
represented by projectors $E_1$ and $E_2$ on particles $\# 1$ and $\#
2$, respectively.

We can now view the two particles as a closed  or as an open
system in which the measuring devices are a part of the environment. 
For the closed system the time dependence of the $E_k$ is 
\be
E_k(t_k)=V_k(t_k)^{\dag} E_k V_k(t_k)=
e^{iH_k t_k}E_k e^{-iH_k t_k}
\ee
with time independent Hamiltonians. 

Directly measurable probabilities are:
\begin{itemize}
\item
Probability of the result ``yes" for $E_1$ on particle $\# 1$
\be
P[E_1(t_1)]=\langle\Psi_0|E_1(t_1)\otimes I_2|\Psi_0\rangle,
\ee
\item
Probability of the result ``yes" for $E_2$ on particle $\# 2$
\be
P[E_2(t_2)]=\langle\Psi_0|I_1\otimes E_2(t_2)|\Psi_0\rangle,
\ee
\item
Joint probability of results ``yes" for both particles
\be
P[E_1(t_1) \cap E_2(t_2)]=
\langle\Psi_0|E_1(t_1)\otimes E_2(t_2)|\Psi_0\rangle.\label{joint}
\ee
\end{itemize}
The conditional probability of the result ``yes" for $E_2$ on particle $\# 2$
under the condition that ``yes" is found for $E_1$ on particle $\# 1$
\be
P[E_2(t_2) | E_1(t_1)]=
\frac{P[E_1(t_1) \cap E_2(t_2)]}{P[E_1(t_1)]},
\ee
is calculated from the joint probability and the probability of the
condition. 

In such an experiment the behaviour of particle $\# 1$ for times later than
$t_1$ is irrelevant. 
The measurement at $t_1$ is a {\it destructive\/} measurement
of the property represented by $E_1(t_1)$. 

In the Heisenberg picture one expects that any operator, also
$E_1\otimes E_2$, has a unitary time dependence
\be
E_1(t)\otimes E_2(t)
&=&
U(t)^{\dag}
E_1\otimes E_2
U(t)
\ee
with some generator $H$. 
For the open system 
we construct a Hamiltonian in which 
the parameters $t_1$ and $t_2$ are encoded. Such an
operator is interesting from the fundamental point of view and is
also essential for later applications. The following time dependent
Hamiltonian has the required properties
\be
H_{t_1,t_2}(t)&=& \theta(t-t_1) H_1\otimes I_2+
\theta(t-t_2) I_1\otimes H_2.\label{kill}
\ee
were $\theta(x)$ is the step function equal 1 for $x<0$ and 0
otherwise (note that $\theta(x)=\Theta(-x)$ where $\Theta$ is the
Heaviside function). 
$t_k$ are parameters indicating the times when
interaction with the detectors takes place. The 
evolution of the projectors is 
\be
E_1(t)\otimes E_2(t)
&=&
e^{i\int_0^t H_{t_1,t_2}(t')dt'}
E_1\otimes E_2
e^{-i\int_0^t H_{t_1,t_2}(t')dt'}.\nonumber
\ee
In particular, 
\be
E_1(t)\otimes E_2(t)
=E_1(t_1)\otimes E_2(t_2)\quad {\rm if}\, t_1,t_2\leq t.
\ee

\section{Correlation experiments in linear quantum mechanics
--- Schr\"odinger picture}

How to do the same calculation in the Schr\"odinger picture? 
This is a relevant question in our context, since in nonlinear
quantum mechanics the Heisenberg picture in the usual sense does not
exist. 

As in Sec. II there are at least two posibilities which are based on a
``projection at-a-distance" (closed system), and the time
dependent Hamiltonian (open system). 

\subsection{Projection-at-a-distance approach}

The dynamics of the state is 
\be
|\Psi(t)\rangle &=& 
V_1(t)\otimes V_2(t)|\Psi_0\rangle.\label{evolution}
\ee
The calculation of the probabilities can be done with the following algorithm.
\begin{itemize}
\item
Evolve the two-particle state until $t=t_1$ by means of (\ref{evolution}).
\item
At $t=t_1$ project with $E_1\otimes I_2$ and normalize
\be
|\Psi(t_1)\rangle\mapsto \frac{E_1\otimes I_2 |\Psi(t_1)\rangle}
{\parallel E_1\otimes I_2 |\Psi(t_1)\rangle\parallel}
=:|\tilde\Psi(t_1)\rangle
\label{step1}.
\ee
The projector $E_1$ represents the proposition which gave the result
``yes" in the measurement performed on particle $\# 1$.
\item
Evolve the resulting state for $t_1<t<t_2$ starting at $t_1$ with the
initial condition (\ref{step1}) by means of 
$I_1\otimes V_2(t-t_1)$, i.e.
\be
I_1\otimes V_2(t-t_1)
|\tilde\Psi(t_1)\rangle.
\label{step2}
\ee
\item
Calculate at $t=t_2$ the average of $I_1\otimes E_2$
in the state (\ref{step2})
\be
\frac{\langle \Psi(t_1)|
E_1\otimes V_2(t_2-t_1)^{\dag}E_{2}V_2(t_2-t_1) |\Psi(t_1)\rangle}
{\langle \Psi(t_1)|E_1\otimes I_2 |\Psi(t_1)\rangle}\label{step3}.
\ee
This is the conditional probability of the result
``yes" for the second particle under the condition that the appropriate
measurement gave ``yes" for the first particle.
\item
The interpretation of the denominator in (\ref{step3}) 
shows that the joint probability 
is given by the numerator of (\ref{step3}), 
\be
\langle\Psi_0|
V_1(t_1)^{\dag}E_{1}V_1(t_1)\otimes
V_2(t_2)^{\dag}E_{2}V_2(t_2) |\Psi_0\rangle.\label{step4}
\ee
which is the formula we wanted to derive.
Here the conditional probability and the probability of the condition
imply the joint probability.
\end{itemize}

\subsection{Open-system approach}

There exists a simpler and more
straightforward method of computing the correlation function if we
use the time dependent Hamiltonian (\ref{kill}).
Solving the SE with (\ref{kill}) we find 
\be
|\Psi_{t_1,t_2}(t)\rangle
=
e^{-i H_1\otimes I_2\int_0^t \theta(\tau-t_1)d\tau   
-iI_1\otimes H_2\int_0^t\theta(\tau-t_2)d\tau}
|\Psi_0\rangle.\label{state12}
\ee
The joint probability (\ref{joint}) is, like in the Heisenberg
picture, directly available,
\be
P[E_1(t_1) \cap E_2(t_2)]=
\langle \Psi_{t_1,t_2}(t)|E_1\otimes E_2|\Psi_{t_1,t_2}(t)\rangle,
\label{joint''}
\ee
with $t_1, t_2\leq t$. 

\section{Nonlinear Hamiltonian evolutions} 

We restrict the nonlinear one-particle
Schr\"odinger equations, for simplicity, to the classical Hamiltonian class,
i.e. to those which can be written as 
\be
i\dot \psi_A(x)
=
\{\psi_A(x),{\cal H}\}
=
\frac{\delta {\cal H}}{\delta \bar \psi^{A}(x)}.
\ee
Linear Schr\"odinger-type equations are in this class; 
furthermore also some nonlinear
Schr\"odinger equations (NLSE) can be formulated in this way
(``$|\psi(x)|^2$ NLSE"
\cite{FT}, the Bia{\l}ynicki-Birula--Mycielski NLSE \cite{BBM},
certain family of Doebner--Goldin NLSE \cite{DG}, and the
equations discussed by Weinberg \cite{Weinberg}).
Weinberg's NLSE simultaneously belong to a family of generalized SE
defined in an analogous way on projective spaces and K\"ahler manifolds
\cite{Kibble,Cirelli,Ashtekar,Hughston,Bona}. 

As mentioned in the introduction an extension of the dynamics from one to many
particles can be constructed, in the tensor product space, 
in different ways. If one wants a local two-particle NLSE (for example, such
that a potential applied to one of the particles does not influence
the other one) the extensions are restricted.
Of particular interest in this context is the sub-class of 
one-particle NLSE with Hamiltonian functions 
satisfying the Polchinski condition
\cite{Polchinski}:
\be
{\cal H}(\psi,\bar\psi)={\cal H}(\rho)\big|_{\rho=|\psi\rangle\langle\psi|}.
\ee
For example, in a two-dimensional Hilbert space 
$|\psi\rangle=\left(\begin{array}{c}\psi_+\\\psi_-\end{array}\right)$
represents a spin-1/2 system. The Hamiltonian function 
\be
{\cal H}(\psi,\bar\psi)
&=&
{\cal H}(\psi_+,\psi_-,\bar\psi_+,\bar\psi_-)
= (\psi_+\bar\psi_-+\psi_-\bar\psi_+)^2\nonumber\\
&=&
\langle
\psi|\sigma_x|\psi\rangle^2
=
(\Tr|\psi\rangle\langle\psi|\sigma_x)^2\nonumber\\
&=&
(\Tr\rho\sigma_x)^2\big|_{\rho=|\psi\rangle\langle\psi|}
=:
{\cal H}(\rho)\big|_{\rho=|\psi\rangle\langle\psi|}
\ee
satisfies the Polchinski condition, whereas 
\be
{\cal H}(\psi,\bar\psi)
&=&
(\psi_+\psi_-+\bar\psi_-\bar\psi_+)^2\label{19}
\ee
does not: (\ref{19}) is not invariant under $|\psi\rangle\mapsto 
e^{i\alpha}|\psi\rangle$. 

In linear quantum mechanics Hamiltonian functions can be
written as 
\be
{\cal H}(\psi,\bar\psi)
&=&
\langle\psi|H|\psi\rangle
=
\Tr\big(|\psi\rangle\langle\psi| H\big)\nonumber\\
&=&
\Tr\rho H\big|_{\rho=|\psi\rangle\langle\psi|}
=:
{\cal H}(\rho)\big|_{\rho=|\psi\rangle\langle\psi|}
\ee
and, hence, fulfil the condition. 
Bia{\l}ynicki-Birula--Mycielski and ``$|\psi(x)|^2$"
are examples of NLSE satisfying the Polchinski condition. A weakened
version of the condition is applicable to all Doebner--Goldin
equations \cite{MC98}.

Assume now that we have two particles with 
Hamiltonian functions fulfilling the above
criterion, i.e. 
\be
{\cal H}_1(\psi_1,\bar\psi_1)
&=&{\cal H}_1(\rho)\big|_{\rho=|\psi_1\rangle\langle\psi_1|},\\
{\cal H}_2(\psi_2,\bar\psi_2)
&=&{\cal H}_2(\rho)\big|_{\rho=|\psi_2\rangle\langle\psi_2|},
\ee
and a generic {\it entangled\/} state \footnote{From now on we 
employ notation more appropriate for systems with discrete degrees of
freedom. This is motivated by finite-dimensional examples we will
discuss later.} 
\be
|\Psi(t)\rangle
&=&
\sum_{k_1k_2}\Psi(t)_{k_1k_2}|k_1\rangle|k_2\rangle.
\ee 
States of 
the one particle subsystems may be
represented by reduced density matrices [$\Psi=\Psi(t)$]
\be
\rho_1
&=&
\sum_{k_1l_1 k_2}\bar \Psi_{k_1k_2}\Psi_{l_1k_2}
|k_1\rangle\langle l_1|,\\ 
\rho_2
&=&
\sum_{k_1 k_2 l_2}\bar \Psi_{k_1k_2}\Psi_{k_1l_2}
|k_2\rangle\langle l_2|.
\ee
Polchinski defined a {\it two-particle\/} Hamiltonian function 
by their sum evaluated at appropriate one-particle states of
particles $\#1$ and $\#2$, respectively, i.e. as 
\be
{\cal H}_{1+2}(\Psi,\bar\Psi)
&:=&{\cal H}_1(\rho)\big|_{\rho_1}
+
{\cal H}_2(\rho)\big|_{\rho_2}. 
\ee
The corresponding two-particle NLSE has the Hamiltonian form 
\be
i\dot \Psi_{k_1k_2} &=& \frac{\partial {\cal H}_{1+2}(\Psi,\bar\Psi)}
{\partial \bar\Psi_{k_1k_2}}.\label{1+2}
\ee
In typical situations (see the Appendix) the solution of (\ref{1+2})
can be written as 
\be
|\Psi(t)\rangle
&=&
V_1(\Psi_0,t)\otimes V_1(\Psi_0,t)|\Psi_0\rangle\\
&=&
V_1(\rho_1(0),t)\otimes V_1(\rho_2(0),t)|\Psi_0\rangle.
\ee
We can write with its help reduced density
matrices of the subsystems. 
It can be shown at different levels of
generality \cite{Polchinski,Jordan,MC97} 
that the dynamics of a reduced density matrix of one of
the subsystems is
independent of the choice of Hamiltonian function of the other
subsystem (for a simple proof see Appendix). This establishes
locality of the extension.

\section{Correlation experiments in nonlinear quantum mechanics
--- Schr\"odinger picture}

We mentioned already that in nonlinear quantum mechanics the usual
Heisenberg picture may not exist. 
For a nonlinear evolution of pure one-particle states the
Schr\"odinger picture is automatically given. Hence we describe the
correlation experiment in the Schr\"odinger picture. We have shown
that there are two posibilities: the projection-at-a-distance
approach and the open-system approach. In the linear case they give
the same results (which agree also with those from the Heisenberg
picture). 

The projection-at-a-distance approach was employed to two-particle
systems in nonlinear quantum mechanics by Gisin \cite{Gisin} and
recently by Mielnik \cite{Mielnik}. 
The conclusion of these papers was that a nonlocal effect necessarily
appears independently of the form of one-particle nonlinearity and
the form of two-particle extension.
In the next section we show on an explicit example and using the
Polchinski extension that the above conclusion is correct if
one sticks to this particular representation of the projection postulate. 
However, the argument does not work if one uses an open-system approach.

To adapt to nonlinear quantum mechanics 
the open-system approach one has to modify the two-particle
extension. We generalize the Polchinski 
two-particle Hamiltonian function as follows
\be
{\cal H}_{t_1,t_2}(t,\Psi,\bar\Psi)
&=&\theta(t-t_1){\cal H}_1(\rho)\big|_{\rho_1}
+
\theta(t-t_2){\cal H}_2(\rho)\big|_{\rho_2}. \label{new open}
\ee
The Schr\"odinger equation for the two particles reads again 
\be
i\dot \Psi_{k_1k_2} &=& 
\frac{\partial {\cal H}_{t_1t_2}(t,\Psi,\bar\Psi)}
{\partial \bar\Psi_{k_1k_2}}\label{K1+2}
\ee
($\Psi_{k_1k_2}=\Psi_{t_1,t_2}(t)_{k_1k_2}$). Solutions of
(\ref{K1+2}) are of the form 
(cf. Sec. VI and the Appendix)
\be
|\Psi_{t_1,t_2}(t)\rangle
&=&
V_1(\rho_1(0),t,t_1)\otimes V_2(\rho_2(0),t,t_2)|\Psi_0\rangle\label{Sol1}
\ee
where $V_k$ depend only on (nonlinear and time dependent) 
Hamiltonians and initial
reduced density matrices of $k$-th particles.

It follows that the reduced density matrices are 
\be
\rho_k(t)
=
V_k(\rho_k(0),t,t_k)
\rho_k(0)
V_k(\rho_k(0),t,t_k)^{\dag},\quad k=1,2.\label{Sol2}
\ee
As a consequence one cannot
influence the dynamics of particle $\# 1$ by modifications of
potentials, moments of detection, and initial conditions
corresponding to particle $\# 2$, and vice versa. 
This establishes locality of the dynamics. 

Let us note that the open-system approach is independent of
projections at-a-distance and one
can directly use the formula from linear quantum mechanics:
If $|\Psi_{t_1,t_2}(t)\rangle$ is a solution
of (\ref{K1+2}) then, for $t_1,t_2\leq t$, the joint probability is
\be
P[E_1(t_1) \cap E_2(t_2)]=
\langle \Psi_{t_1,t_2}(t)|E_1\otimes E_2|\Psi_{t_1,t_2}(t)\rangle.
\label{joint'''}
\ee
To illustrate how this works we consider an
explicit example.

\section{Example: Evolution of a pair of spin-1/2 particles}
\label{sec-ex}

We start with one-particle Hamiltonian functions
\be
{\cal H}_1(\rho)
&=&
A[\Tr(\rho \sigma_z)]^2/2\\
{\cal H}_2(\rho)
&=&
B[\Tr(\rho \sigma_z)]^2/2\\
{\cal H}_1(\psi_1,\bar\psi_1)
&=&
A[\Tr(|\psi_1\rangle\langle\psi_1|\sigma_z)]^2/2\\
&=&
A\langle \psi_1|\sigma_z|\psi_1\rangle^2/2\\
{\cal H}_2(\psi_2,\bar\psi_2)
&=&
B[\Tr(|\psi_2\rangle\langle\psi_2|\sigma_z)]^2/2\\
&=&
B\langle \psi_2|\sigma_z|\psi_2\rangle^2/2.
\ee
$A$ and $B$ are real constants and $|\psi_1\rangle$, $|\psi_2\rangle$
are one-particle state-vectors. The corresponding one-particle
equations obtained from these Hamiltonian finctions are
\be
i|\dot\psi_1\rangle
&=&
A\langle \psi_1|\sigma_z|\psi_1\rangle \sigma_z
|\psi_1\rangle,\\
i|\dot\psi_2\rangle
&=&
B\langle \psi_2|\sigma_z|\psi_2\rangle \sigma_z
|\psi_2\rangle.
\ee
Both nonlinear Hamiltonian operators are of the form 
$
\bbox b\cdot \bbox \sigma
$
where 
\be
\bbox b\sim (0,0,\langle \sigma_z\rangle).
\ee
This is a mean-field type interaction of a Curie-Weiss type.

The Polchinski two-particle extension is
\be
{\cal H}_{1+2}(\Psi,\bar\Psi)
&=&
A[\Tr(\rho_1 \sigma_z)]^2/2
+
B[\Tr(\rho_2 \sigma_z)]^2/2\\
&=&
A\langle\Psi|\sigma_z\otimes I|\Psi\rangle^2/2
+
B\langle\Psi|I\otimes \sigma_z|\Psi\rangle^2/2
\ee
and the two-particle Schr\"odinger equation derived from this
Hamiltonian function is 
\be
i|\dot \Psi\rangle
&=&
\Big(
A\langle\Psi|\sigma_z\otimes I|\Psi\rangle
\sigma_z\otimes I
+
B\langle\Psi|I\otimes \sigma_z|\Psi\rangle
I\otimes \sigma_z
\Big)
|\Psi\rangle.\label{42}
\ee
\subsection{Open-system approach}

The generalized Polchinski two-particle Hamiltonian function is
\be
{\cal H}_{t_1,t_2}(\Psi,\bar\Psi)
&=&
\theta(t-t_1)A\langle\Psi|\sigma_z\otimes I|\Psi\rangle^2/2
+
\theta(t-t_2)B\langle\Psi|I\otimes \sigma_z|\Psi\rangle^2/2
\ee
and
\be 
i|\dot \Psi\rangle
&=&
\Big(
\theta(t-t_1)A\langle\Psi|\sigma_z\otimes I|\Psi\rangle
\sigma_z\otimes I
+
\theta(t-t_2)B\langle\Psi|I\otimes \sigma_z|\Psi\rangle
I\otimes \sigma_z
\Big)
|\Psi\rangle.\label{43}
\ee
The general solution of (\ref{43}) is
\be
|\Psi_{t_1,t_2}(t)\rangle
&=&
e^{-i A\langle\Psi_0|\sigma_z\otimes I|\Psi_0\rangle
\sigma_z\otimes I \int_0^t\theta(\tau-t_1)d\tau
-i
B\langle\Psi_0|I\otimes \sigma_z|\Psi_0\rangle
I\otimes \sigma_z
\int_0^t\theta(\tau-t_2)d\tau}
|\Psi_0\rangle\nonumber\\
&=&
e^{-i A \langle\sigma_z(0)\rangle_1\sigma_z 
\kappa(t,t_1)}
\otimes
e^{-iB \langle\sigma_z(0)\rangle_2
\sigma_z 
\kappa(t,t_2)}
|\Psi_0\rangle\label{gen-sol}
\ee
where $\langle\sigma_z(0)\rangle_k=\Tr(\rho_k(0)\sigma_z)$, 
$\kappa(t,t_k)=\int_0^t\theta(\tau-t_k)d\tau$. The averages in the
exponents are evaluated in $|\Psi_0\rangle$. This is a consequence
of 
\be
\langle\Psi_0|\sigma_z\otimes I|\Psi_0\rangle
&=&
\langle\Psi_{t_1,t_2}(t)|\sigma_z\otimes I|\Psi_{t_1,t_2}(t)\rangle
\\
\langle\Psi_0|I\otimes \sigma_z|\Psi_0\rangle
&=&
\langle\Psi_{t_1,t_2}(t)|I\otimes \sigma_z|\Psi_{t_1,t_2}(t)\rangle
\ee
as one can verify by direct substitution. 

(\ref{gen-sol}) describes the entire history of the two particles:
From their ``birth" at $t=0$ to their ``deaths" at $t=t_1$ and
$t=t_2$. The solution of (\ref{42}) is recovered in the limits
$t_1,t_2\to +\infty$. 

Using (\ref{gen-sol}) we can explicitly compute the dynamics of the
two subsystems. The reduced density matrices are 
\be
\rho_1(t)
&=&
e^{-i A \langle\sigma_z(0)\rangle_1\sigma_z 
\kappa(t,t_1)}
\rho_1(0)
e^{i A \langle\sigma_z(0)\rangle_1\sigma_z 
\kappa(t,t_1)}\\
\rho_2(t)
&=&
e^{-iB \langle\sigma_z(0)\rangle_2
\sigma_z 
\kappa(t,t_2)}
\rho_2(0)
e^{iB \langle\sigma_z(0)\rangle_2
\sigma_z 
\kappa(t,t_2)}.
\ee
The form of the above explicit solutions is instructive because of 
the following properties:
\begin{itemize}
\item
The subsystems evolve independently of each other. 
\item
The solutions are uniquely determined by the initial condition
$|\Psi_0\rangle$ at $t=0$.
\item
The evolution operator for the pair is 
\be
V_1(\Psi_0,t)\otimes V_2(\Psi_0,t)
=
V_1(\rho_1(0),t)\otimes V_2(\rho_2(0),t)
\ee
i.e. 
is a product of {\it unitary\/} operators which 
depend on $\rho_k(0)$ and not on their decompositions in
particular bases. 
\end{itemize}
From the solution (\ref{gen-sol}) one can calculate correlation
functions for any observable (see Sec. V).

Operationally there is no ambiguity in the open-system formulation.
If one wants to know predictions for an experiment one has to insert the
detection times, $t_1$ and $t_2$, into (\ref{gen-sol}).  

In an actual experiment one deals with $N$ pairs. If we assume for simplicity
that for all the pairs the times of flight $\Delta t_k^i=t_k^i-t_0^i$,
$k=1,2$, $i=1,\dots,N$, are the same and equal $\Delta t_k$ 
we can compute averages of observables, say, $X_1\otimes X_2$,
by
\be
\langle X_1\otimes X_2\rangle_{\Psi,\Delta t_1,\Delta t_2}
=
\langle \Psi_{\Delta t_1,\Delta t_2}(t)|X_1\otimes X_2
|\Psi_{\Delta t_1,\Delta t_2}(t)\rangle.
\ee
Averages of one-system observables, say $X_1$, are computed in the standard
way 
\be
\langle X_1\otimes I_2\rangle_{\Psi,\Delta t_1,\Delta t_2}
&=&
\langle \Psi_{\Delta t_1,\Delta t_2}(t)|X_1\otimes I_2
|\Psi_{\Delta t_1,\Delta t_2}(t)\rangle\nonumber\\
&=&
\Tr\Big(
e^{i A\langle \sigma_z(0)\rangle_1
\sigma_z\kappa(t,\Delta t_1)}
X_1e^{-i A\langle \sigma_z(0)\rangle_2
\sigma_z\kappa(t,\Delta t_1)}
\rho_1(0)\Big).
\ee
The average does not depend on $\Delta t_2$. As we have
already said this is a consequence of the local properties of the
Polchinski extension. 

\subsection{Projection-at-a-distance approach}

We follow the calculation from Sec. III A step by step. 
Consider measurements of spin in direction $\bbox a_k$, $k=1,2$, 
i.e. the observable is $X_k=\bbox a_k\cdot \bbox\sigma$ 
with projectors $E_k=E_k^\pm=(I_k\pm X_k)/2$. 
\begin{itemize}
\item
At $t=t_1$ the state is
\be
|\Psi(t_1)\rangle
&=&
\underbrace{
e^{-i A \langle\sigma_z(0)\rangle_1\sigma_z  t_1}
}_{V_1(\Psi_0,t_1)}
\otimes
\underbrace{
e^{-iB \langle\sigma_z(0)\rangle_2
\sigma_z  t_1}}_{V_2(\Psi_0,t_1)}
|\Psi_0\rangle.
\ee
\item
At $t=t_1$ project with $ E_1^\pm\otimes I_2$ and normalize
\be
|\Psi(t_1)\rangle
\mapsto
\frac{E_1^\pm\otimes I_2 |\Psi(t_1)\rangle}
{\parallel E_1^\pm\otimes I_2 |\Psi(t_1)\rangle \parallel}
=:|\Psi_\pm(t_1)\rangle.\label{psi_pm}
\ee
\item
Evolve the resulting state for $t_1<t<t_2$ 
but starting at $t_1$ with the initial condition (\ref{psi_pm})
\be
|\Psi_\pm(t_2)\rangle
&=&
I_1
\otimes
\underbrace{
e^{-iB \langle\Psi_\pm(t_1)|I_1\otimes \sigma_z|\Psi_\pm(t_1)\rangle
\sigma_z  (t_2-t_1)}}_{V_2\big(\Psi_\pm(t_1),t_2-t_1\big)}
|\Psi_\pm(t_1)\rangle.
\ee
\item
Compute the conditional probability 
\be
{}&{}&\langle\Psi_\pm(t_2)|
I_1\otimes E_2^s
|\Psi_\pm(t_2)\rangle\nonumber\\
&{}&\pp =
=
\frac{\langle \Psi(t_1)|
E_1^\pm
\otimes
e^{iB \langle\Psi_\pm(t_1)|I_1\otimes \sigma_z|\Psi_\pm(t_1)\rangle
\sigma_z  (t_2-t_1)}
E_2^{s}e^{-iB \langle\Psi_\pm(t_1)|I_1\otimes \sigma_z|\Psi_\pm(t_1)\rangle
\sigma_z  (t_2-t_1)}
|\Psi(t_1)\rangle}
{\langle \Psi(t_1)|E_1^\pm \otimes I_2 |\Psi(t_1)\rangle}\label{condit'}
\ee
where $E_2^s$ is $E_2^+$ or $E_2^-$.
\end{itemize}
The joint probability 
\be
\langle \tilde\Psi^\pm_{t_1,t_2}(t_2)|
E_1^\pm
\otimes
E_2^{s}
|\tilde\Psi^\pm_{t_1,t_2}(t_2)\rangle,
\ee
where 
\be
|\tilde\Psi^\pm_{t_1,t_2}(t_2)\rangle
=
e^{-iA \langle\Psi_0|\sigma_z\otimes I|\Psi_0\rangle \sigma_z t_1}
\otimes
e^{-iB \langle\Psi_\pm(t_1)|I\otimes \sigma_z|\Psi_\pm(t_1)\rangle
\sigma_z  (t_2-t_1)}
e^{-iB \langle\Psi_0|I\otimes \sigma_z|\Psi_0\rangle \sigma_z t_1}
|\Psi_0\rangle,
\ee
can be calculated from 
(\ref{condit'}). 

Just for comparison let us note that 
the open-system calculation produces at this point joint probability
of the form 
\be
\langle \Psi_{t_1,t_2}(t_2)|
E_1^\pm
\otimes
E_2^{s}
|\Psi_{t_1,t_2}(t_2)\rangle.
\ee
Now we can pinpoint the difference between the two approaches. 
The frequencies of spin rotation are different. In the
projection-at-a-distance approach we have
$$
B \langle\Psi_\pm(t_1)|I_1\otimes \sigma_z|\Psi_\pm(t_1)\rangle
$$
and in the open-system approach
$$
B \langle\Psi_0|I\otimes \sigma_z|\Psi_0\rangle.
$$
They depend on the projected state taken at $t_1$ and the initial
state at $t=0$, respectively. 

\subsection{Numerical example}

For a numerical illustration of previous considerations we take
a convenient initial state 
\be
|\Psi_0\rangle
&=&
\frac{1}{3}|1\rangle|2\rangle
-
\frac{2\sqrt{2}}{3}|2\rangle|1\rangle
\ee
where 
\be
|1\rangle
=
\left(
\begin{array}{c}
\cos(\pi/8)\\
\sin(\pi/8)
\end{array}
\right),
\quad
|2\rangle
=
\left(
\begin{array}{c}
-\sin(\pi/8)\\
\cos(\pi/8)
\end{array}
\right).
\ee
The parameters in Hamiltonians are $A=8$, $B=1/2$, and the detection
times are $t_1=3.5$ and $t_2=8$ (all in dimensionless units). 
Figs. 1 and 2 show averages of 
$\sigma_x \otimes \sigma_x$ (solid), 
$\sigma_x\otimes I_2$ (dashed), and 
$I_1\otimes \sigma_x$ (dotted)
calculated by means of the two 
approaches. 


In Fig. 1 we used the open-system approach. The dotted
line representing the average of $I_1\otimes \sigma_x$ does not
``notice" the measurement performed on particle $\#1$. In Fig. 2 the
projection at-a-distance was employed. 
One can observe a slight change in the doted curve at $t=t_1$. This
is the nonlocal effect of the type described by Gisin \cite{Gisin}. 
Until $t=t_1$
the evolution is described in the open-system way. One can see
from the figures that 
projection-at-a-distance reasoning leads even in this case to the
nonlocal influence between the two particles. 

\section{Nonlinear generalization of projection at-a-distance}

As the final step of our analysis we show that there exists a
generalization of the projection-at-a-distance algorithm leading to
results equivalent to those from the open-system approach. 
The algorithm is applicable if there is no causal relation between
the correlated measurements. Modifications are needed if causal
realtions do occur (see the next section). 
The modified algorithm follows steps analogous to those from Sec. VI:
\begin{itemize}
\item
Evolve the two-particle state until $t=t_1$ by means of the evolution
generated by (\ref{new open}). The solution has the form (see
Appendix) 
\be
|\Psi_{t_1,t_2}(t)\rangle
&=&
V_1(\Psi_0,t)\otimes V_2(\Psi_0,t)|\Psi_0\rangle.\label{aa-sol}
\ee
\item
At $t=t_1$ project and again normalize
\be
|\Psi_{t_1,t_2}(t_1)\rangle\mapsto \frac{E_1\otimes I_2 
|\Psi_{t_1,t_2}(t_1)\rangle}
{\parallel E_1\otimes I_2 |\Psi_{t_1,t_2}(t_1)\rangle\parallel}\label{step1'}.
\ee
\item
Evolve this state by  
\noindent
$I_1\otimes V_2(\Psi_0,t-t_1)$, i.e.
\be
\frac{E_1\otimes V_2(\Psi_0,t_2-t_1)
|\Psi_{t_1,t_2}(t_1)\rangle} 
{\parallel E_1\otimes I_2 |\Psi_{t_1,t_2}(t_1)\rangle\parallel}\label{step2'}
\ee
\item
Calculate at $t=t_2$ the average of $I_1\otimes E_2$
in the state (\ref{step2'})
\be
\frac{\langle \Psi_{t_1,t_2}(t_1)|
E_1\otimes V_2(\Psi_0,t_2-t_1)^{\dag}E_{2}V_2(\Psi_0,t_2-t_1)
|\Psi_{t_1,t_2}(t_1)\rangle}
{\langle \Psi_{t_1,t_2}(t_1)|E_1\otimes I_2 
|\Psi_{t_1,t_2}(t_1)\rangle}\label{step3'}.
\ee
\end{itemize} 
The denominator in (\ref{step3'}) is the probability of the condition.
Therefore 
the joint probability 
is given by the numerator of (\ref{step3'}). Using (\ref{aa-sol}) we
obtain (\ref{joint'''}). 

As we can see there is only one modification with respect to the
derivation which led in the example to the nonlocal effect:
Instead of 
\be
V_2\big(\Psi_\pm(t_1),t_2-t_1\big)\label{oldV}
\ee
the following expression appears
\be
V_2(\Psi_0,t_2-t_1)=V_2(\rho_2(0),t_2-t_1),
\ee
where $|\Psi_0\rangle$ and $\rho_2(0)$ are the {\it initial\/} conditions for
the pair and the second particle, respectively. 

\section{Remarks on causally related correlation experiments}

Our discussion was purposefully restricted to measurements which
are spacelike separated. 
However, a dual problem remains: What about measurements
which are {\it not\/} spacelike separated, a
situation one encounters in preparation at-a-distance? 

\subsection{Preparation at-a-distance and teleportation}

Preparation at-a-distance is a procedure which produces a state of a
physical system $\# 1$ on the basis of destructive measurements
performed on a correlated system $\# 2$. Active quantum teleportation
is a particular case of this procedure.
The procedure is often referred to as a {\it non-destructive\/} measurement.  

Assume, for example, that we have to produce ``an ensemble of white stones"
which are selected at random from a box containing black and white
pebbles. How do we do this? We take a randomly chosen stone and 
``look at it". If the stone is white, we keep it. Otherwise we thow
it away. 

The experimental setup involves two steps. In the first step we
scatter some light on the stones and our eyes perform {\it
destructive\/} measurements of the 
scattered photons. The second step involves a {\it local action\/}
(keeping or removing the stone) performed on the ensemble of black and white
pebbles. 

The second step is as necessary for the preparation as the first
one, and is performed in the {\it future light-cone\/} of the
detection event. In 
(active) quantum teleportation an analogue of the second step 
is typically referred to as a ``classical communication
channel supplemented by local operations". 
We cannot prepare in such a way an ensemble of white stones (or spins
``up") in a
region of space-time which is spacelike separated from the detection
area. For the same reason teleportation cannot be faster than light. 

In the next section we discuss a
probabilistic game which in many respects is analogous
to the nonlinear EPR problem. The example shows that in correlation
experiments involving a nonlinear dynamics one has to take into
account propagation of information between correlated subsystems. 

\subsection{Russian roulette with a cheating player}

The nonlinear EPR problem is not, in its essence, a problem of quantum
mechanics. It is a general difficulty present in all nonlinear systems whose
dynamics depends on probability and which involve reduction of
probability via correlations.
The Russian roulette with a cheating player is an example of a situation
where the required properties occur.

The Russian roulette is a game whose simplest version is the following. 
There are two players, Anna and Boris, a gun with two chambers, and one
bullet. 
The players do not know which chamber is loaded.
They put certain amounts of money into the pool and 
Boris begins the
game: He points the gun at himself and pulls the trigger.
If he is unlucky then Anna wins and
collects all the money. 
However, if the bullet was in the other chamber, the next move belongs to
Anna...

There are two variants of continuation. 

(a) Anna is not informed about the result of Boris' trial before she
pulls the trigger.

(b) She knows what happened to Boris. 

The first case involves separated events. 
The nontrivial formal element of the game is the behaviour of Anna
in the second case. If Boris had a bad day she can safely pull the trigger and
wins. In the opposite case she knows this time the gun will fire and 
it makes no sense to continue, so she tries to cheat. 

How to formally model the game? Thinking of a {\it real-life\/}
version of the duel it is rather clear that the change of behaviour of a
player is due to his (her) lack of knowledge about the actual
location of the cartridge in the gun. Real versions of the game
involved six players and a six-chamber gun. As the game continues the
probability that the next player will get killed increases if the
players are not allowed to randomly spin the chambers after each
trial. It seems that in a formal model of the game we should assume
that the behaviour of the cheating player (who nevertheless tries to
spin the chamber) is probability dependent (the greater the
probability of getting killed the greater the motivation to
circumvent the rules). 
If we agree on this viewpoint the roulette becomes an
interesting playground for testing the concepts of probability 
reduction in systems whose dynamics is {\it probability dependent\/}. 

More instructive is the version of the game with a six-chamber gun
and three bullets which are placed in such a way that between two
loaded chambers there is an empty one. 
Each time one pulls the trigger the chamber shifts by one place. 
If Boris was lucky then the loaded chamber is in place and she
should cheat (rotate it by one position). However, if Boris was not lucky
she will shoot herself if she cheats. Cheating and non cheating
are here statistically equivalent: One half of the ensemble of Annas
will not survive the game if they are not informed, independently of
whether they cheat or not. 

We can say that the dynamics of Anna is
independent of probability (i.e. linear) if she is not informed.
Let us note that this is exactly analogous to the example
we give in Sec.~\ref{sec-ex}: The nonlinearity vanishes if the
average involves the entire density matrix of the subsystem. 

The behaviour of Anna changes at the moment she gets the information
and not at the moment Boris makes his ``measurement". This can be
verified statistically since now the entire ensemble of cheating Annas will
survive. This variant is analogous to the EPR problem as discussed by
Gisin in \cite{Gisin}: The nonlinearity reacts to the density matrix
which involves reduced probabilities. 
Still, the reduction of Anna's probability is not instantaneous.
How to
describe the reduction is a completely different issue. An
interesting discussion of a similar problem can be found in a recent paper
by Kent \cite{Kent}. 

Let us finally note that the link of the game to the nonlinear EPR problem
becomes even more evident if one assumes that Anna makes her decision
on the basis of an {\it incomplete\/} information. Then in order to
survive she estimates the probability that the information she
obtained is reliable and her behaviour is explicitly probability
dependent. 

\section{Summary}

Among other obstructions for the formulation of a physically motivated and
mathematically decent nonlinear extension of quantum mechanics one
encounters the following problem: How to build from a one-particle
system a time evolution of a multi-particle one, and how to compute
correlation experiments in this system. There is an additional
condition: We want a local theory. Hence we use the Polchinski
multi-particle extension which is sufficient for a local description
of equal-time correlation experiments. To include multiple-time and
spacelike-separated 
correlation experiments we generalize the Polchinski formalism by
treating the system as an open one with detectors in the role of an
environment. Now multi-particle Hamiltonians are time dependent and
parametrized by the detection times. On this basis we derive a
generalization of the projection-at-a-distance algorithm which is
appropriate for nonlinear correlation experiments with spacelike
separated events. The modified algorithm predicts 
the same probabilities as the open-system generalization of the
Polchinski approach and the nonlocal effects are eliminated. 
We also give a new argument against an instantaneous reduction of
probability in correlation experiments.

\acknowledgments

MC thanks Alexander von Humboldt Foundation for making possible his
stay in Clausthal where this work was done, Polish Committee for
Scientific Research for support by means of the KBN Grant No. 5 PO3B 040 20,
and Ania and Wojtek Pytel for the notebook computer used to type-in
this paper. Our understanding of the problem was influenced by
numerous discussions we had with Nicolas Gisin, Gerald Goldin, Adrian
Kent, Wolfgang L\"ucke, and Bogdan Mielnik. 

\section{Appendix: Solutions of (31) and locality}

Take Hamiltonian functions 
\be
{\cal H}_1(\rho_1)
&=&
{\cal H}_1(\{\rho_{1a_1b_1}\})
=
{\cal H}_1(\{\sum_{c_2}\Psi_{a_1c_2}\bar\Psi_{b_1c_2}\})\\
{\cal H}_2(\rho_2)
&=&
{\cal H}_1(\{\rho_{2a_2b_2}\})
=
{\cal H}_1(\{\sum_{c_1}\Psi_{c_1a_2}\bar\Psi_{c_1b_2}\}).
\ee
Define the following Hamiltonian operators 
\be
H_1(\rho_1)_{b_1a_1}
&=&
\frac{\partial {\cal H}_1}{\partial \rho_{1a_1b_1}}
\\
H_2(\rho_2)_{b_2a_2}
&=&
\frac{\partial {\cal H}_2}{\partial \rho_{2a_2b_2}}.
\ee
The operators are Hermitian since $\rho_1$ and $\rho_2$ are
Hermitian. Let ${\cal H}_{t_1,t_2}$ be given by (\ref{new open}).
Using the chain rule 
one can show that 
\be
i|\dot\Psi\rangle
&=&
\sum_{k_1k_2}
\frac{\partial {\cal H}_{t_1,t_2}}{\partial \bar\Psi_{k_1k_2}}
|k_1\rangle|k_2\rangle\nonumber\\
&=&
\Big(
\theta(t-t_1)
H_1(\rho_1)\otimes I_2
+
\theta(t-t_2)
I_1\otimes H_2(\rho_2)
|\Psi\rangle.\label{cauchy}
\ee
If the Cauchy problem for (\ref{cauchy}) is well posed, its solution
$|\Psi(t)\rangle$ is 
uniquely determined by the initial condition $|\Psi_0\rangle$ at
$t=0$. Assume $|\Psi(t)\rangle=|\Psi[\Psi_0,t]\rangle$ is known. 
Substituting the solution into (\ref{cauchy}) and denoting 
\be
\tilde H_k(\Psi_0,t)
&=&
\theta(t-t_k)
H_k\big(\rho_k(\Psi[\Psi_0,t],\bar\Psi[\Psi_0,t])\big)\\
&=&
\theta(t-t_k)
H_k\big(\rho_k(t)\big)\\
&=&
\tilde H_k(\rho_k(0),t)
\ee
we can see that $|\Psi(t)\rangle$ is a solution of 
\be
i|\dot\Psi_{t_1,t_2}(t)\rangle
&=&
\Big(
\tilde H_1(\Psi_0,t)\otimes I_2
+
I_1\otimes \tilde H_2(\Psi_0,t)
\Big)
|\Psi_{t_1,t_2}(t)\rangle.\label{cauchy'}
\ee
For a fixed initial value $|\Psi_{t_1,t_2}(0)\rangle
=|\Psi_0\rangle$ this is a {\it linear\/} Schr\"odinger equation with
time-dependent Hamiltonian (the dependence on the set of parameters
defining the initial condition
is nonlinear). 
Using results from linear quantum mechanics we conclude
that there exist unitary operators $V_k(\Psi_0,t)=V_k(\rho_k(0),t)$ 
such that 
\be
|\Psi(t)\rangle
&=&
V_1(\Psi_0,t)\otimes V_2(\Psi_0,t)|\Psi_0\rangle\\
&=&
V_1(\rho_1(0),t)\otimes V_2(\rho_2(0),t)|\Psi_0\rangle.\label{a-sol}
\ee
To each $|\Psi_0\rangle$ there corresponds an orbit of the dynamics. 
The difference with respect to linear
quantum mechanics is that on different orbits we have different
unitary evolutions. 

The reduced density matrices evolve by 
\be
\rho_k(t)
=
V_k(\rho_k(0),t)
\rho_k(0)
V_k(\rho_k(0),t)^{\dag}.
\ee
The behaviour of the subsystems is determined entirely by local
Hamiltonians and local initial conditions for states. This
establishes locality. 

This would not be the case if $\tilde H_k(\Psi_0,t)$ did not depend
on one-particle states of the $k$th particle. This also shows that
different local two-particle extensions may be possible if 
different one-particle representations of states are used.

\twocolumn

\begin{figure}
\epsfxsize=8.25cm
\epsffile{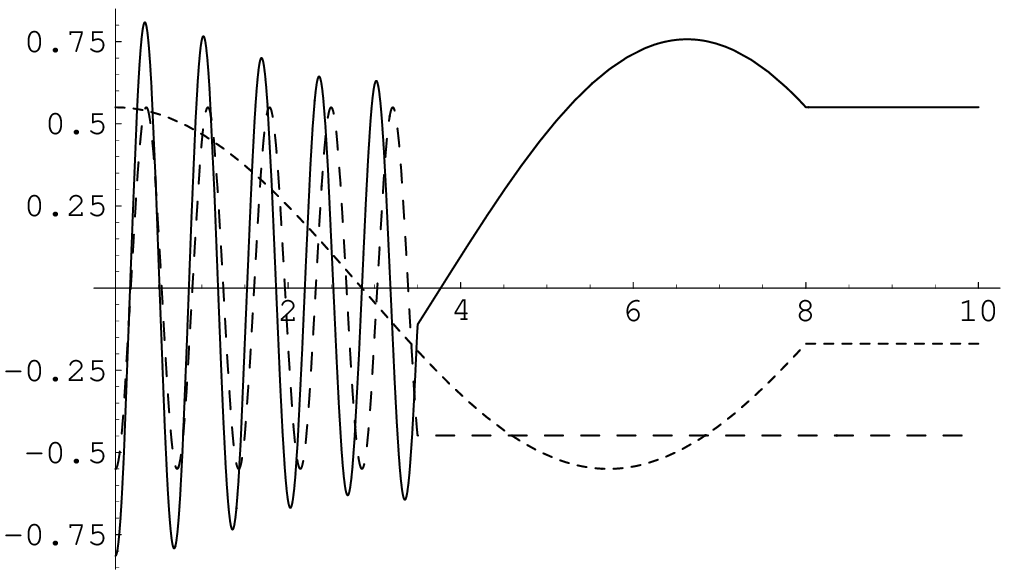}
\caption{Averages of the three observables in the open-system
formulation. The dotted line shows the evolution of observable
$\sigma_x$ associated with particle $\#2$ which is detected at
$t=t_2=8$. Earlier detection of particle $\#1$ at $t_1=3.5$ does not
influence particle $\#2$.} 
\end{figure}
\begin{figure}
\epsfxsize=8.25cm
\epsffile{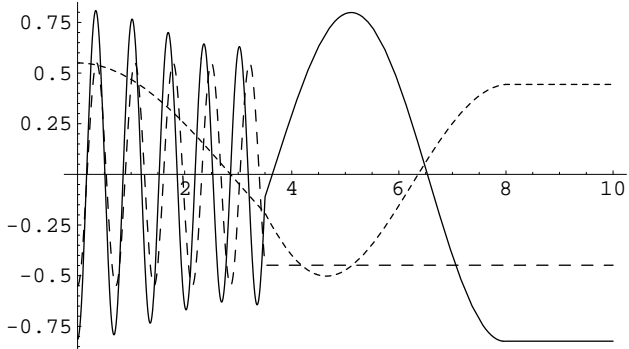}
\caption{Averages of the three observables in the
standard projection-at-a-distance formulation. 
Measurement at $t=t_1=3.5$ performed on particle $\#1$
nonlocally influences the behaviour of particle $\#2$. As opposed to
the plot from Fig. 1 the dotted line is modified at $t=3.5$. This is
Gisin-type nonlocality.} 
\end{figure}
\end{document}